\documentclass[11pt,a4paper]{article}
\pdfoutput=1

\usepackage{enumitem}
\usepackage{etex}
\usepackage{amsthm}
\usepackage{geometry}
\usepackage{dcolumn}
\usepackage{epsf}
\usepackage{mathrsfs}
\usepackage{multirow}
\usepackage{booktabs}
\usepackage{tabularx}
\usepackage{array}
\usepackage{slashed}
\usepackage{float}
\usepackage{leftidx}
\usepackage{setspace}
\usepackage{verbatim}
\usepackage{adjustbox}
\usepackage{tikz}
\usepackage[caption=false]{subfig}
\usepackage{arydshln}
\usepackage{jheppub}

\numberwithin{equation}{section}
\usetikzlibrary{arrows,decorations.markings,shapes.arrows,patterns,positioning}

\newcommand{\bea}{\begin{eqnarray}}
\newcommand{\eea}{\end{eqnarray}}
\newcommand{\bean}{\begin{eqnarray*}}
\newcommand{\eean}{\end{eqnarray*}}

\def\W #1{\widetilde{#1}}

\def\Label#1{\label{#1}%
  \smash{\hbox to0pt{\raise1ex\hbox{\tiny[#1]}\hss}}}
\def\Label#1{\label{#1}}
\renewcommand{\eqref}[1]{eq.~(\ref{#1})}

\newcommand{\figref}[1]{figure~\ref{#1}}

\newcommand{\secref}[1]{section~\ref{#1}}

\def\braket#1{\left\langle #1 \right\rangle}

\def\vev{\braket}

\def\bvev#1{\left[ #1 \right]}
\def\Spaa{\vev}
\def\Spbb{\bvev}

\def\rchi{\raisebox{\depth}{$\chi$}}

\newcommand{\ctobedelete}[1]{}

\allowdisplaybreaks
\hypersetup{pageanchor=false}
\title{CHY formula and MHV amplitudes}
\author[a]{Yi-Jian Du}
\emailAdd{yijian.du@whu.edu.cn}
\author[c]{Fei Teng}
\emailAdd{Fei.Teng@utah.edu}
\author[b,c]{Yong-Shi Wu}
\emailAdd{wu@physics.utah.edu}
\affiliation[a]{Center for Theoretical Physics,
School of Physics and Technology,
Wuhan University,\\
299 Bayi Road, Wuhan 430072,
China}
\affiliation[b]{Department of Physics and
Center for Field Theory and Particle Physics, Fudan University,\\
220 Handan Road, Shanghai 200433, China}
\affiliation[c]{Department of Physics and Astronomy, University of Utah,\\ 115 South 1400 East, Salt Lake City, UT 84112, USA}

\abstract{In this paper, we study the relation between the Cachazo-He-Yuan (CHY) formula and the maximal-helicity-violating (MHV) amplitudes of Yang-Mills and gravity in four dimensions. We prove that only one special rational solution of the scattering equations found by Weinzierl supports the MHV amplitudes. Namely, localized at this solution, the integrated CHY formula produces the Parke-Taylor formula for MHV Yang-Mills amplitudes as well as the Hodges formula for MHV  gravitational amplitudes, with an {\em arbitrary} number of external gluons/gravitons. This is achieved by developing techniques, in a manifestly M\"obius covariant formalism, to explicitly compute relevant reduced Pfaffians/determinants. We observe and prove two interesting properties (or identities), which facilitate the computations. We also check that all the other $(n-3)!-1$ solutions to the scattering equations do not support the MHV amplitudes, and prove analytically that this is indeed true for the other special rational solution proposed by Weinzierl, that actually supports the anti-MHV amplitudes. Our results reveal a mysterious feature of the CHY formalism that in Yang-Mills and gravity theory, solutions of scattering equations, involving only external momenta, somehow know about the configuration of external polarizations of the scattering amplitudes.         
 }
\begin{document}
\maketitle
\hypersetup{pageanchor=true}
\section{Introduction} 

Scattering-equation-based formula \cite{Cachazo:2013gna,Cachazo:2013hca,Cachazo:2013iea} proposed by Cachazo, He and Yuan (CHY) provides a new perspective for understanding scattering amplitudes in relativistic quantum field theories.
%
The CHY 
formalism was proved in \cite{Dolan:2013isa} by using Britto-Cachazo-Feng-Witten (BCFW) recursion \cite{Britto:2004ap, Britto:2005fq}. Up to now a lot of efforts have been made on understanding the CHY formula, including generalization to various theories \cite{Cachazo:2014nsa,Cachazo:2014xea,Naculich:2014naa, Naculich:2015zha,Naculich:2015coa,delaCruz:2015raa}, the study on the scattering equations and their solutions \cite{Monteiro:2013rya,Kalousios:2013eca,Weinzierl:2014vwa,Cardona:2015ouc,Huang:2015yka,Dolan:2015iln,Sogaard:2015dba,Cardona:2015eba}, new soft theorems from the CHY formula \cite{Cachazo:2015ksa},  off-shell extension \cite{Lam:2015mgu},  the relationship to Feynman diagrams \cite{Cachazo:2015nwa,Baadsgaard:2015ifa}, discussions on worldsheet theories \cite{Mason:2013sva,Bjerrum-Bohr:2014qwa} and generalizations to loop level \cite{Geyer:2015bja,Adamo:2013tsa,Casali:2014hfa,Adamo:2015hoa,Ohmori:2015sha,Baadsgaard:2015hia,He:2015yua,Geyer:2015jch,Cachazo:2015aol,Feng:2016nrf}.

However, although the CHY formalism 
is highly compact, it is really hard to obtain explicitly the analytic results expressed in terms of Lorentz invariant variables (for example, $s_{ab}$) for scattering amplitudes. In a sense, this is because neither the solutions to the scattering equations nor the relation between the CHY formula and Feynman diagrams is easily available. {Among the efforts on solving scattering equations, (as far as we know) solutions in four dimensions are studied first in the work \cite{Monteiro:2013rya}. Shortly after, Weinzierl proposed two special rational solutions in four dimensions in terms of spinor variables \cite{Weinzierl:2014vwa} (for details, see \eqref{eq:sl1} and (\ref{eq:sl2}) and the discussion there):
\bea
\sigma_{a}^{(1)}&=&\frac{\langle {a},{n-2}\rangle\langle {n-1},{\rchi}\rangle}{\langle {a},{\rchi}\rangle\langle {n-1},{n-2}\rangle}\,,\Label{eq:sl1a}\\
\sigma_{a}^{(2)}&=&\frac{ [{a},{n-2}] [{n-1},{\rchi}]}{ [{a},{\rchi}] [{n-1},{n-2}]}\,,\Label{eq:sl2a}
\eea
which were conjectured to correspond to MHV and anti-MHV amplitudes in \cite{Monteiro:2013rya} and \cite{Naculich:2014naa}.} There has been no explicit proof of the statement that the integrated CHY formula of these two solutions exactly reproduce the famous Parke-Taylor formula \cite{Parke:1986gb,Xu:1986xb} for MHV (and anti-MHV) amplitudes. On the other hand, though the relationship between Feynman rules and CHY integrations was already established for scalar amplitudes \cite{Cachazo:2015nwa,Baadsgaard:2015ifa}, it has not been able to derive the Parke-Taylor formula for generic MHV (and anti-MHV) tree-level Yang-Mills amplitudes following this line of thoughts. Moreover, it is not apparent at all to see that the Hodges formula \cite{Hodges:2012ym} for gravity MHV (and anti-MHV) amplitudes at tree level is also supported by these two solutions.

In this paper, we fill the gap by explicitly demonstrating that the special solution (\ref{eq:sl1a}) supports the Parke-Taylor formula for MHV Yang-Mills amplitudes as well as the Hodges formula for MHV gravitational amplitudes, with an {\em arbitrary} number of external gluons/gravitons. Similarly, the solution (\ref{eq:sl2a}) supports the anti-MHV amplitudes for Yang-Mills and gravity. To show this, our proof proceeds as follows:
\begin{itemize}
    \item The original integrated CHY formula expresses amplitudes by summing over terms localized at different solutions to the scattering equations. We first consider only the term contributed by the special rational solution (\ref{eq:sl1a}), and the reduced Pfaffian of $\Psi$ for a fixed-helicity MHV configuration $(1^{-},2^{-},3^{+},\ldots,n^{+})$, from which we can prove that the Parke-Taylor formula for the color-ordered MHV Yang-Mills amplitude, as well as the Hodges formula for MHV gravitational amplitude, are reproduced. Two interesting properties of the reduced Pfaffian make the proof tractable:
    \begin{itemize}
        \item {\bf \textit{Property-1}} The reduced Pfaffian of $\Psi$ at the MHV configuration can be expanded in terms of determinants of reduced $C$ matrices with three columns and three rows deleted. This property relies on the MHV configuration, but is independent of which solution we choose.
        \item {\bf \textit{Property-2}} Both the reduced Pfaffian of $\Psi$ and the reduced determinant of $\Phi$ localized at the solution (\ref{eq:sl1a}) can be expressed in terms of the Hodges formula for gravitational amplitude.
    \end{itemize}
    \item We then extend our discussion to general color-ordered MHV amplitudes (with the two negative helicities at arbitrary positions). This is achieved by extending the two properties to more general cases, which can also be understood by considering the Kleiss-Kuijf (KK) relation \cite{Kleiss:1988ne}.

    \item Finally, one needs to check that any solution other than \eqref{eq:sl1a} leads to a zero reduced determinant of $C$ (with three rows and colums deleted as in Property-2). For the other Weinzierl solution (\ref{eq:sl2a}), this can be proved analytically.
 \end{itemize}
 There are two interesting observations in this approach that deserve more attention:
\begin{itemize}
\item {In property-2, the building blocks ${\det}'(\Phi)$ and $\text{Pf}\,'(\Phi)$ are expressed in terms of the gravitational amplitude $\bar{M}_n(12\ldots n)$, while only the pre-factors in front of it can be changed by an $SL(2,\mathbb{C})$ transformation.} Thus the $SL(2,\mathbb{C})$ invariance of both the color-ordered Yang-Mills MHV amplitude and the gravity MHV amplitude becomes manifest.
\item The vanishing of the reduced determinant of $C$ actually imposes constraints on solutions to the scattering equations. With these constraints, one can distinguish the solution (\ref{eq:sl1a}) contributing to MHV amplitudes from the other $(n-3)!-1$ solutions. Similar statement is also true for the solution (\ref{eq:sl2a}) that supports anti-MHV amplitudes. We hope that such classification can be extended to other solutions.
\end{itemize}

This paper is organized as follows. Section \ref{Sec2} presents a review of the CHY formula, Parke-Taylor formula and Hodges formula, and defines our notations. In \secref{Sec3}, we prove that the special rational solution given by Weinzierl reproduces the Parke-Taylor formula for MHV Yang-Mills amplitudes and the Hodges formula for MHV gravitational amplitudes. This is achieved by a M\"obius covariant calculation. In \secref{Sec4}, we check that other solutions do not contribute to the MHV configuration. Especially, this is analytically proved for the other Weinzierl rational solution (\ref{eq:sl2a}). We then propose a set of complex polynomial equations that distinguishes the special solution (\ref{eq:sl1a}) that supports MHV amplitudes from the others. Finally, we devote \secref{Sec5} to a summary of our results and discussions on possible extensions. Some useful properties of the spinor helicity formalism and details of the proof are given in appendix \ref{appA} and \ref{appB} respectively.

\section{Preparation: a review of the CHY, Parke-Taylor and Hodges formula}
\label{Sec2}
In this section, we present a warm-up review of some useful details of the general CHY formula (\ref{eq:CHY}) for scattering amplitudes, the Parke-Taylor formula (\ref{Parke-Taylor-xy}) for MHV Yang-Mills amplitudes, as well as the Hodges formula (\ref{Eq:HodgesFormula}) for MHV gravitational amplitudes, all in four dimensions.

\subsection{CHY formula}
CHY proposed in a series of papers \cite{Cachazo:2013gna,Cachazo:2013hca,Cachazo:2013iea} that any $n$-point tree amplitude $A_n(1,2,\dots,n)$ in arbitrary dimensions can be expressed by the following equation:
\begin{equation}
A_n(1,2,\dots,n)=\int\frac{dz_{1}\ldots dz_{n}}{\text{Vol}\left[SL(2,\mathbb{C})\right]}{\prod_{a}}\,'\delta\left(\sum_{b\neq a}\frac{s_{ab}}{z_{ab}}\right)\mathcal{I}_{n}\,. \Label{eq:CHY}
\end{equation}
The building blocks of \eqref{eq:CHY} are discussed in the following:
\begin{itemize}[]
\item{\bf Scattering equations}

The scattering equations for $n$ massless particles, which are imposed by delta functions in \eqref{eq:CHY}, are
\begin{equation}
\sum_{b\neq a}\frac{s_{ab}}{z_{ab}}=0\,,\quad a\in\{1,2,\ldots,n\}\,,\Label{eq:se}
\end{equation}
where $s_{ab}=2k_{a}\cdot k_{b}$ and $z_{ab}\equiv z_{a}-z_{b}$. The scattering equation is $SL(2,\mathbb{C})$ covariant, namely, if the set $\{\sigma_{a}\}$ is a solution, the set $\{\zeta_{a}\}$ with
\begin{align}
\label{sl2c}
    &\zeta_{a}=\frac{\alpha\sigma_{a}+\beta}{\gamma\sigma_{a}+\delta}\,,& &\alpha,\beta,\gamma,\delta\in\mathbb{C}\,,& &\alpha\delta-\beta\gamma=1\,,&
\end{align}
is also a solution. We can thus use this freedom to fix three arbitrarily chosen $z$'s to three arbitrary positions on the Riemann sphere, say, $(z_{p},z_{q},z_{r})=(\sigma_{p},\sigma_{q},\sigma_{r})$. The first consequence is that there are only $n-3$ independent equations in (\ref{eq:se}). It can be proved by a semi-analytical inductive method that the number of solutions is $(n-3)!$ in any dimension \cite{Cachazo:2013gna}. The second consequence is that the integration over $z_{p}$, $z_{q}$ and $z_{r}$ actually encodes the $SL(2,\mathbb{C})$ redundancy. Using a Fadeev-Popov like trick, we can divide out the volume of the $SL(2,\mathbb{C})$ group in \eqref{eq:CHY}:
\begin{equation}
    \frac{dz_{1}\ldots dz_{n}}{\text{Vol}\left[SL(2,\mathbb{C})\right]}=\prod_{c\neq p,q,r}dz_{c}\left(\sigma_{pq}\sigma_{qr}\sigma_{rp}\right)\,.
\end{equation}
\item{\bf The integrated CHY formula}

In \eqref{eq:CHY}, after integrating the $z$ variables over the permutation invariant delta-function
\begin{equation}
\label{deltafun}
    {\prod_{a}}\,'\delta\left(\sum_{b\neq a}\frac{s_{ab}}{z_{ab}}\right)\equiv\text{perm}(ijk)\,\sigma_{ij}\sigma_{jk}\sigma_{ki}\prod_{a\neq i,j,k}\delta\left(\sum_{b\neq a}\frac{s_{ab}}{z_{ab}}\right)\,,
\end{equation}
the scattering amplitudes can be expressed by the following form
\begin{equation}
    A_{n}=\text{perm}(ijk)\,\text{perm}(pqr)\sum_{\{\sigma\}\in\text{solutions}}\frac{\sigma_{pq}\sigma_{qr}\sigma_{rp}\sigma_{ij}\sigma_{jk}\sigma_{ki}}{\det\left(\Phi_{ {p}, {q}, {r}}^{ {i}, {j}, {k}}\right)}\,\mathcal{I}_{n}\,,\Label{CHY-integrated}
\end{equation}
where the factor $\text{perm}(pqr)$ is the signature of the permutation that moves the standard ordering $(1,2,\ldots n)$ to the ordering $(p,q,r,\ldots)$, with $(\ldots)$ always keeping the ascending order. If both $(ijk)$ and $(pqr)$ are in the ascending order, we have
\begin{equation*}
    \text{perm}(ijk)\,\text{perm}(pqr)=(-1)^{i+j+k+p+q+r}.
\end{equation*}
$\Phi$ is an $n\times n$
 matrix given by
\begin{equation}
\renewcommand{\arraystretch}{1.5}
\Phi_{ab}=\left\{\begin{array}{>{\displaystyle}l @{\hspace{1.5em}} >{\displaystyle}l}
\frac{s_{ab}}{\sigma_{ab}^{2}} & a\neq b \\
-\sum_{c\neq a}\frac{s_{ac}}{\sigma_{ac}^{2}} & a=b \\
\end{array}\right.\,,\Label{Phi}
\end{equation}
and
$\left(\Phi_{ {p}, {q}, {r}}^{ {i}, {j}, {k}}\right)$ is the matrix obtained by removing the $(i,j,k)$-th row and the $(p,q,r)$-th column from $\Phi$. Its determinant, $\det\left(\Phi^{i,j,k}_{p,q,r}\right)$, is nothing but the Jacobian associated with the delta-functions in \eqref{deltafun}. If we define the reduced determinant ${\det}'(\Phi)$ to be
\begin{equation}
   {\det}'(\Phi)\equiv\text{perm}(ijk)\,\text{perm}(pqr)\frac{\det\left(\Phi_{pqr}^{ijk}\right)}{\sigma_{ij}\sigma_{jk}\sigma_{ki}\sigma_{pq}\sigma_{qr}\sigma_{rp}}\,,~~\Label{eq:rd}
\end{equation}
the amplitude  (\ref{CHY-integrated}) can then be expressed simply as
\begin{equation}
    A_{n}=\sum_{\{\sigma\}\in\text{solutions}}\frac{\mathcal{I}_{n}}{{\det}'(\Phi)}\,.
\end{equation}
This form suggests that to compute the amplitudes, we need to know all the solutions to the scattering equations and sum over the contributions from all of them. 

\item{\bf The integrand for gauge theory and gravity}

Finally, the integrand $\mathcal{I}_{n}$ for color-ordered gauge amplitudes is set to
\bea
\label{kernel-YM}
    \mathcal{I}_{n}(\{k,\epsilon,\sigma\})=\frac{\text{Pf}\,'(\Psi)}{\sigma_{12}\sigma_{23}\ldots \sigma_{n1}}\,,
\eea
while for gravitational amplitudes, we use
\begin{equation}
\label{kernel-gravity}
     \mathcal{I}_{n}\left(\{k,\epsilon,\W{\epsilon},\sigma\}\right)={\text{Pf}\,'\left[\Psi(k,\epsilon,\sigma)\right]\times\text{Pf}\,'\left[\Psi(k,\W{\epsilon},\sigma)\right]}\,,
\end{equation}
where $\epsilon$ and $\widetilde{\epsilon}$ together give the polarizations of the external gravitons. The reduced Pfaffian in \eqref{kernel-YM} and \eqref{kernel-gravity} is proportional to the Pfaffian of $\Psi$ with both the $(i,j)$-th row and $(i,j)$-th column removed:
\begin{equation}
\text{Pf}\,'(\Psi)=\frac{\text{perm}(ij)}{\sigma_{ij}}\text{Pf}(\Psi_{ {i} {j}}^{ {i} {j}})\,,\Label{eq:rf}
\end{equation}
where $1\leq i<j\leq n$. Here the $2n\times 2n$ antisymmetric matrix $\Psi$ is given by
\begin{equation}
\Psi(\{k,\epsilon,\sigma\})=\left(\begin{array}{cc}
A & -C^{T} \\
C & B \\
\end{array}\right),~~\Label{Eq:Psi}
\end{equation}
where $k$, $\epsilon$ denote the momenta and polarization vectors of external particles. The matrices $A$, $B$ and $C$ are defined by
\begin{align}
& A_{ab}=\left\{\begin{array}{>{\displaystyle}l @{\hspace{1em}} >{\displaystyle}l}
\frac{s_{ab}}{\sigma_{ab}} & a\neq b\\
0 & a=b \\
\end{array}\right.\,,&
& B_{ab}=\left\{\begin{array}{>{\displaystyle}l @{\hspace{1em}} >{\displaystyle}l}
\frac{2\epsilon_{a}\cdot\epsilon_{b}}{\sigma_{ab}} & a\neq b\\
0 & a=b \\
\end{array}\right.\,,&
& C_{ab}=\left\{\begin{array}{>{\displaystyle}l @{\hspace{1em}} >{\displaystyle}l}
\frac{2\epsilon_{a}\cdot k_{b}}{\sigma_{ab}} & a\neq b\\
-\sum_{c\neq a}\frac{2\epsilon_{a}\cdot k_{c}}{\sigma_{ac}} & a=b \\
\end{array}\right.\,.
\label{Eq:ABCMatrices}
\end{align}

\item{\bf Two rational solutions in four dimensions}

In four dimensions, one can express light-like 4-vectors in term of spinors. Two of the solutions to the scattering equations have been found to be rational functions of spinor variables \cite{Weinzierl:2014vwa}:
\bea
\sigma_{a}^{(1)}&=&\frac{\langle {a},{n-2}\rangle\langle {n-1},{\rchi}\rangle}{\langle {a},{\rchi}\rangle\langle {n-1},{n-2}\rangle}\,,\Label{eq:sl1}\\
\sigma_{a}^{(2)}&=&\frac{ [{a},{n-2}] [{n-1},{\rchi}]}{ [{a},{\rchi}] [{n-1},{n-2}]}\,.\Label{eq:sl2}
\eea
The others solutions are expected to be more complicated algebraic functions of spinor variables. The spinor convention we have adopted in this work is given in appendix \ref{appA}. When writing down this solution, we have implicitly fixed part of the $SL(2,\mathbb{C})$ freedom by choosing
\begin{align*}
    &\sigma_{n-2}=0\,,& &\sigma_{n-1}=1\,,
\end{align*}
for all the solutions. The arbitrary spinor $|\rchi\rangle$ represents the remaining $SL(2,\mathbb{C})$ freedom, and we are going use a formalism that is manifestly covariant under this freedom. It is not difficult to generalize to a formalism that is totally M\"obius covariant, which is mentioned in \secref{mobius}. In the following sections, we will show that the relevant quantities like reduced Pfaffian/determinant are of a factorized form, in which the $\chi$-dependent and $\chi$-independent factors can be separately identified. It turns out that in the final expressions for physical MHV amplitudes, the $\chi$-dependent factors, which represents part of the $SL(2,\mathbb{C})$ freedom, are all canceled, making the invariance under this freedom manifest. Thus, despite the appearance of the $\chi$-dependence in the intermediate steps, our calculation is actually $SL(2,\mathbb{C})$ covariant, once properly generalized. If we set $|\rchi\rangle=|n\rangle$, we will return to the original form presented in \cite{Weinzierl:2014vwa} and have $\sigma_{n}=\infty$. For the two solutions in \eqref{eq:sl1} and \eqref{eq:sl2}, we have the following expressions for $\sigma_{ab}=\sigma_{a}-\sigma_{b}$:
\begin{align}
&\sigma_{ab}^{(1)}=\frac{\langle {a},{b}\rangle\langle {n-2},{\rchi}\rangle\langle {n-1},{\rchi}\rangle}{\langle {a},{\rchi}\rangle\langle {b},{\rchi}\rangle\langle {n-1},{n-2}\rangle}\,,& &\sigma_{ab}^{(2)}=\frac{ [{a},{b}] [{n-2},{\rchi}] [{n-1},{\rchi}]}{ [{a},{\rchi}] [{b},{\rchi}] [{n-1},{n-2}]}\,, ~~\Label{Eq:Sigmaij}
\end{align}
which will be used frequently later.
\end{itemize}

{\subsection{Parke-Taylor formula}
The color-ordered tree-level Yang-Mills MHV amplitude $A^{\text{MHV}}_n$ with two negative-helicity gluons $x$ and $y$ ($1\leq x < y\leq n$) is given by Parke-Taylor formula \cite{Parke:1986gb,Xu:1986xb}
\bea
A^{\text{MHV}}_n(1^+,\dots,x^-,\dots,y^-,\dots,n^+)=\frac{\langle xy\rangle^{4}}{\langle 12\rangle\langle 23\rangle\ldots\langle n1\rangle}\,.~~\Label{Parke-Taylor-xy}
\eea
Replacing $\Spaa{\ldots}$ by $\Spbb{\ldots}$, one can obtain the anti-MHV amplitudes.

\subsection{Hodges formula}

The gravitational reduced MHV superamplitude  $\bar{M}_n(12\ldots n)$ for the $N=7$ formulation\footnote{As stated in e.g., \cite{Elvang:2013cua}} of $N=8$ supergravity} can be expressed by the Hodges formula \cite{Hodges:2012ym}
\begin{equation}
    \bar{M}_n(12\ldots n)=(-1)^{n+1}\text{perm}(ijk)\,\text{perm}(pqr)c_{ijk}c^{pqr}\det\left(\phi_{p,q,r}^{i,j,k}\right)\,,\Label{Eq:HodgesFormula}
\end{equation}
where the $c$ symbol is
\begin{equation}
    c_{abc}=c^{abc}=\frac{1}{\langle ab\rangle\langle bc\rangle\langle ca\rangle}\,.
\end{equation}
Here, we use the notation of \cite{Cachazo:2013hca}, in which $(i,j,k)$ denotes the deleted rows and $(p,q,r)$ denotes the deleted columns, while \cite{Hodges:2012ym} uses the opposite convention. The $\phi$ matrix is define by
\begin{align}
    &\phi_{ab}=\frac{[ab]}{\langle ab\rangle}\,,& &\phi_{aa}=-\sum_{\substack{l=1\\l\neq a}}^{n}\frac{[al]\langle l\rchi\rangle\langle l1\rangle}{\langle al\rangle\langle a\rchi\rangle\langle a1\rangle}\,,\Label{Eq:HodgesMatrix}
\end{align}
for $2\leq a\neq b\leq n$. We note that $\phi_{aa}$ is invariant if we change the spinor $|1\rangle$ or $|\rchi\rangle$ into any $|\theta\rangle$ that is not collinear with $|a\rangle$. As an example, for $a\neq n$, we multiply $\langle an\rangle$ into both the numerator and denominator
\begin{align}
\label{gaugeinv}
    \phi_{aa}&=-\sum_{\substack{l=1\\l\neq a}}^{n}\frac{[al]\langle l\rchi\rangle\langle l1\rangle\langle an\rangle}{\langle al\rangle\langle a\rchi\rangle\langle a1\rangle\langle an\rangle}\nonumber\\
    &=-\sum_{\substack{l=1\\l\neq a}}^{n}\left(\frac{[al]\langle l1\rangle\langle n\rchi\rangle}{\langle a\rchi\rangle\langle a1\rangle\langle an\rangle}+\frac{[al]\langle ln\rangle\langle l1\rangle}{\langle al\rangle\langle an\rangle\langle a1\rangle}\right)=-\sum_{\substack{l=1\\l\neq a}}^{n}\frac{[al]\langle ln\rangle\langle l1\rangle}{\langle al\rangle\langle an\rangle\langle a1\rangle}\,.
\end{align}
It can also be shown that $\bar{M}_n(12\ldots n)$ is independent of any choice of $(i,j,k)$ and $(p,q,r)$ \cite{Hodges:2012ym}. {Using Hodges formula, one can write down the $n$-point MHV gravitational amplitude ${\cal M}_{n}$ immediately
\begin{equation}
    {\cal M}_{n}(1^+,\dots,x^-,\dots,y^-,\dots,n^+)=\Spaa{xy}^8\bar{M}_n(12\ldots n)\,,\Label{Eq:pure-graviton-amp}
\end{equation}
where only the gravitons $x$ and $y$ have negative helicity.

\section{MHV Yang-Mills and gravity amplitudes from CHY formula}
\label{Sec3}
Having prepared the useful properties of the CHY formula for this paper, let us first consider the relation between the CHY formula and the Parke-Taylor formula of MHV amplitudes in four dimensions.
Without loss of generality, we start with the color-ordered MHV amplitude $A^{\text{MHV}}_n\left(1^-,2^-,3^+,\dots,n^+\right)$ where $1$ and $2$ are the two negative helicity gluons.
The Parke-Taylor formula for this amplitude is given by
\bea
A^{\text{MHV}}_n\left(1^-,2^-,3^+,\dots,n^+\right)={\frac{\Spaa{12}^4}{ \Spaa{12}\Spaa{23}\dots\Spaa{n1}}}\,.\Label{Parke-Taylor}
\eea
To relate the Parke-Taylor formula (\ref{Parke-Taylor}) with the CHY formula in four dimensions, we should write the external polarizations by the spinor-helicity formalism \cite{Xu:1986xb}. In appendix \ref{appA}, we have also included a short review of this formalism. It has been shown that the reduced Pfaffian is independent of the gauge choice (namely, the Ward identity holds). In spinor-helicity formalism, one can choose reference momentum for each external gluon to fix the gauge. For the MHV configuration, we can choose the momentum $k_n$ of the gluon $n$ as the reference momentum of the two negative helicity gluons $1$ and $2$. The reference momentum of positive helicity gluons $3,\dots,n$ is chosen as $k_1$. Thus the polarizations of our external gluons are written as
\begin{align}
    &\epsilon_{i}^{\mu}(-)=\frac{\langle i|\gamma^{\mu}|n]}{\sqrt{2}[ni]}\quad(i=1,2)\,,&
    &\epsilon_{j}^{\mu}(+)=\frac{\langle 1|\gamma^{\mu}|j]}{\sqrt{2}\langle 1j\rangle}\quad(j=3,\dots,n)\,.\Label{Eq:Gauge}
\end{align}
In this section, we are going to prove that using only the rational solution given in \eqref{eq:sl1}, we can derive both Parke-Tylor formula and Hodges formula. We first substitute the external polarizations (\ref{Eq:Gauge}) into the integrated CHY formula (\ref{CHY-integrated}) and then show in detail (in appendix \ref{appB}) that the Parke-Taylor formula (\ref{Parke-Taylor}) can really emerges with only the rational solution given in \eqref{eq:sl1}. Then we sketch the calculation that generic MHV amplitudes with negative helicity gluons being at arbitrary positions can also emerge from \eqref{eq:sl1}. In both cases, we  find that $\text{Pf}\,'(\Psi)$ and ${\det}\,'(\Phi)$ are proportional to the reduced gravitational amplitude $\bar{M}_n(12\ldots n)$ defined in \eqref{Eq:HodgesFormula}, which makes the derivation of the MHV gravity amplitude using the CHY formula very straightforward.

Before we start, we need to clarify some terminology. If we say, for example, row-$(i)$ of part $C$, we mean the $i$-th row of the \textit{original} matrix $C$. This is convenient since we constantly delete rows and columns and as a result it is cumbersome to track the position of a specific row in the new matrix after several such manipulations.

\subsection{$A_{n}(1^{-}2^{-}\ldots)$ and $\mathcal{M}_{n}(1^{-}2^{-}\ldots)$ from the integrated CHY formula with solution (\ref{eq:sl1}) }
Now let us prove that the Parke-Taylor formula (\ref{Parke-Taylor-xy}) with $x=1,y=2$ is reproduced by the CHY integral (\ref{CHY-integrated}) localized at the rational solution (\ref{eq:sl1}). Before presenting our proof, we first list two interesting intermediate results:
\begin{enumerate}[label=\bf \textit{Property-{\arabic*}},align=left]
\item The reduced Pfaffian  $\text{Pf}\,'(\Psi)$ in MHV configuration can be expressed by the following expansion in terms of the determinant of $C^{1,2,m}_{1,n-1,n}$ matrices
\begin{equation}
    \text{Pf}\,'(\Psi)={(-1)^{s(n)}}\sum_{m=3}^{n}(-1)^{m}B_{2m}C_{11}\left[\frac{-1}{\sigma_{n-1,n}}\det\left(C^{1,2,m}_{1,n-1,n}\right)\right],~~\Label{Property-1}
\end{equation}
where $C_{p,q,r}^{i,j,k}$ is the matrix $C$ with the row-$(i,j,k)$ and column-$(p,q,r)$ deleted, and the overall sign is controlled by
\begin{equation*}
s(n)=\frac{(n-2)(n-3)}{2}+n+1\,.
\end{equation*}
This property relies on the MHV configuration but is independent of the solutions for the scattering equations.

\item Once we substitute in \eqref{eq:sl1}, $\text{Pf}\,'(\Psi)$, ${\det}\,'(\Phi)$ and the MHV-like factor $\sigma_{12}\ldots\sigma_{n1}$ will have the following compact forms
\begin{subequations}
\label{Property-2}
\begin{align}
    \label{rdPhi}
    {\det}\,'(\Phi)&=\left(F_{\chi}\right)^{2n}\left(P_{\chi}\right)^{4}\bar{M}_{n}(12\cdots n)\,,\\
    \label{MHVfactor}
    \sigma_{12}\ldots\sigma_{n1}&=\left(\frac{1}{F_{\chi}}\right)^{n}\frac{D_{n}}{(P_{\chi})^{2}}\,,\\
    \label{reducedPf}
    \text{Pf}\,'(\Psi)&=(-1)^{s(n)}(\sqrt{2})^{n}\left(F_{\chi}\right)^{n}\left(P_{\chi}\right)^{2}\langle 12\rangle^{4}\bar{M}_{n}(12\cdots n)\,.
\end{align}
\end{subequations}
\end{enumerate}
In \eqref{Property-2}, $\bar{M}_n(12\ldots n)$ is  given by the Hodges formula (\ref{Eq:HodgesFormula}).
 $D_{n}$ is just the denominator of the Parke-Taylor formula
\begin{equation}
    D_{n}=\langle 12\rangle\langle 23\rangle\ldots\langle n1\rangle\,.~~\Label{PTden}
\end{equation}
Finally, the symbol $F_{\chi}$ and $P_{\chi}$ are defined as
\begin{align}
\label{Fn}
    &F_{\chi}\equiv\frac{\langle n-1,n-2\rangle}{\langle n-2,\rchi\rangle\langle n-1,\rchi\rangle}\,,\\
\label{dPn}
    &P_{\chi}\equiv\prod_{a=1}^{n}\langle a\rchi\rangle\,.
\end{align}
Given \eqref{Property-2}, we can see clearly that the CHY formula with the special solution (\ref{eq:sl1}) can repreduce the correct  Parke-Taylor formula (\ref{Parke-Taylor-xy}) for the Yang-Mills MHV amplitude (with a trivial overall factor):
\begin{align}
    \frac{\text{Pf}\,'(\Psi)}{{\det}\,'(\Phi)\times[\sigma_{12}\ldots\sigma_{n1}]}&=(-1)^{s(n)}(\sqrt{2}\,)^{n}\frac{\Spaa{12}^{4}}{\Spaa{12}\Spaa{23}\ldots\Spaa{n1}}\nonumber\\
    &=(-1)^{s(n)}(\sqrt{2}\,)^{n}A_{n}^{\text{MHV}}(1^{-}2^{-}3^{+}\ldots n^{+})\,,
\end{align}
and the Hodges formula (\ref{Eq:pure-graviton-amp}) for the gravitational MHV amplitude:
\begin{equation}
    \frac{\text{Pf}\,'(\Psi)\times\text{Pf}\,'(\Psi)}{{\det}'(\Phi)}=2^{n}\langle 12\rangle^{8}\bar{M}_{n}(12\cdots n)=2^{n}\mathcal{M}_{n}(1^{-}2^{-}3^{+}\ldots n^{+})\,.
\end{equation}

\subsection{General MHV amplitudes}
In this part, we write down the generalized \eqref{Property-1} and \eqref{Property-2} for general MHV amplitudes, in which the negative helicity particles can occupy arbitrary positions. If particles at position $x$ and $y$ ($x<y$ such that $1\leq x\leq n-1$) have negative helicity while all the others are positive, we have
\begin{align}
\label{generalp1}
    \text{Pf}\,'(\Psi)&=(-1)^{s(n)}\sum_{\substack{m=1\\m\neq y}}^{n}(-1)^{y+m+\theta(y-m)+\theta
    (x-m)}B_{ym}C_{xx}\left[\frac{(-1)^{j+n+\theta(x-j)}}{\sigma_{jn}}\det\left(C^{x,y,m}_{x,j,n}\right)\right]\nonumber\\
    &=(-1)^{s(n)}\sum_{\substack{m=1\\m\neq y}}^{n}\text{perm}(xym)\text{perm}(xjn)B_{ym}C_{xx}\left[\frac{1}{\sigma_{jn}}\det\left(C^{x,y,m}_{x,j,n}\right)\right],
\end{align}
where $j$ can be any number except for $x$ and $n$, and the final result does not depend on this choice. In this calculation, we use the polarization vectors
\begin{align}
\label{xygauge}
    \epsilon_{i}^{\mu}(-)=\frac{\langle i|\gamma^{\mu}|n]}{\sqrt{2}[ni]}\quad(i=x,y)\,,&
    &\epsilon_{j}^{\mu}(+)=\frac{\langle x|\gamma^{\mu}|j]}{\sqrt{2}\langle xj\rangle}\quad(1\leq j\neq x,y\leq n)\,.
\end{align}
If we plug the special solution (\ref{eq:sl1}) in, we get
\begin{equation}
\label{generalp2}
    \text{Pf}\,'(\Psi)=(-1)^{s(n)}(\sqrt{2})^{n}(F_{\chi})^{n}(P_{\chi})^{2}\langle xy\rangle^{4}\bar{M}_{n}(12\cdots n)\,.
\end{equation}
On the other hand, ${\det}'(\Phi)$ and $[\sigma_{12}\ldots\sigma_{n1}]$ remain the same as in \eqref{Property-2} since they depend only on kinematics but not helicity configurations. It is thus straightforward to see that the CHY formula gives the desired general MHV Yang-Mills and gravitational amplitudes
\begin{subequations}
\label{generalamp}
\begin{align}
    \frac{\text{Pf}\,'(\Psi)}{{\det}'(\Phi)\times[\sigma_{12}\ldots\sigma_{n1}]}&=(-1)^{s(n)}(\sqrt{2})^{n}\frac{\langle xy\rangle^{4}}{\langle 12\rangle\langle 23\rangle\ldots\langle n1\rangle}\,,\\
    \frac{\text{Pf}\,'(\Psi)\times\text{Pf}\,'(\Psi)}{{\det}'(\Phi)}&=2^{n}\langle xy\rangle^{8}\bar{M}_n(12\ldots n)\,.
\end{align}
\end{subequations}
The derivation of \eqref{generalp1} and \eqref{generalp2} follows closely to those elaborated in \secref{sec:p1} and \ref{sec:p2}, except that one needs to be more careful on the order of the indices involved. The fact that the special rational solution (\ref{eq:sl1}) also gives support to general MHV Yang-Mills amplitudes with arbitrary two negative helicity gluons can also be seen directly from the KK relations for color-ordered Yang-Mills amplitudes \cite{Kleiss:1988ne} and the Parke-Taylor like factors under a given solution \cite{Kol:2014yua}. Similarly, by using only \eqref{eq:sl2}, we can get general anti-MHV amplitudes. All we need to do is to exchange all the angular spinor brackets with the corresponding square spinor brackets, and vice versa.

\subsection{Manifest M\"obius invariance}\label{mobius}
The CHY formalism is of course invariant under M\"obius transformations, namely, if we apply the same $SL(2,\mathbb{C})$ transformation, say, \eqref{sl2c}, to all the solutions, we should still obtain the same physical amplitude. It is thus interesting to explore how the $SL(2,\mathbb{C})$ dependence of our $\text{Pf}\,'(\Psi)$, ${\det}'(\Phi)$ and $[\sigma_{12}\ldots\sigma_{n1}]$ cancel against each other.

Suppose now we put $\sigma_{n}$ to the infinity and use instead the solution proposed by Weinzierl \cite{Weinzierl:2014vwa}:
\begin{equation}
\label{wa}
    w_{a}=\frac{\langle a,n-2\rangle\langle n-1,n\rangle}{\langle a,n\rangle\langle n-1,n-2\rangle}\,,
\end{equation}
then $w_{a}$ and \eqref{eq:sl1} are related through the M\"obius transformation
\begin{equation}
    w_{a}=\frac{\langle n-1,n\rangle\langle n-2,\rchi\rangle \sigma_{a}}{\langle n,\rchi\rangle\langle n-1,n-2\rangle \sigma_{a}-\langle n,n-2\rangle\langle n-1,\rchi\rangle}\,.
\end{equation}
In particular, this transformation leaves $w_{n-2}=\sigma_{n-2}$ and $w_{n-1}=\sigma_{n-1}$, but set $w_{n}=\infty$. Following the CHY formalism, we find that the solution (\ref{wa}) gives
\begin{subequations}
\begin{align}
    {\det}\,'(\Phi)&=\left(F_{n}\right)^{2n}\left(P_{n}\right)^{4}\bar{M}_{n}(12\cdots n)\,,\\
    \sigma_{12}\ldots\sigma_{n1}&=\left(\frac{1}{F_{n}}\right)^{n}\frac{D_{n}}{(P_{n})^{2}}\,,\\
    \text{Pf}\,'(\Psi)&=(-1)^{s(n)}(\sqrt{2})^{n}\left(F_{n}\right)^{n}\left(P_{n}\right)^{2}\langle xy\rangle^{4}\bar{M}_{n}(12\cdots n)\,.
\end{align}
\end{subequations}
In these quantities, the effect of our $SL(2,\mathbb{C})$ transformation is entirely encoded in
\begin{align}
    &F_{\chi}\rightarrow F_{n}=\frac{\langle n-1,n-2\rangle}{\langle n-1,n\rangle\langle n-1,n\rangle}\,,& &P_{\chi}\rightarrow P_{n}=\prod_{a=1}^{n}\langle an\rangle\sim\frac{1}{w_{n}}\prod_{a=1}^{n-1}\langle an\rangle\,.
\end{align}
Clearly $F_{n}$ and $P_{n}$ will cancel each other when calculating physical amplitudes as in \eqref{generalamp}. Although we need a regulator $w_{n}$ for intermediate steps, it does not show up in the physical amplitudes. Now it is also clear that if we want to embrace the full $SL(2,\mathbb{C})$ freedom by relaxing $\sigma_{n-2}$ and $\sigma_{n-1}$ from $0$ and $1$, we should change the spinor $|n-1\rangle$ and $|n-2\rangle$ in the solution (\ref{eq:sl1}) by some other spinors, say $|\theta\rangle$ and $|\eta\rangle$. Then in the final result, we only need to make the replacement
\begin{equation}
    F_{\chi}=\frac{\langle n-1,n-2\rangle}{\langle n-2,\rchi\rangle\langle n-1,\rchi\rangle}\;\rightarrow\;\frac{\langle \theta\eta\rangle}{\langle \eta\chi\rangle\langle\theta\chi\rangle}\,.
\end{equation}
Here the arbitrary choice in the spinors $|\theta\rangle$, $|\eta\rangle$ and $|\chi\rangle$ represents the full $SL(2,\mathbb{C})$ freedom. Thus our calculation explicitly verifies that the MHV amplitudes resulting from the integrated CHY formula is invariant under the M\"obius transformations acting on the solution (\ref{eq:sl1}). Moreover, we have shown that the $SL(2,\mathbb{C})$ dependent pieces factorize out of the gauge invariant building blocks of physical amplitudes, and they cancel with each other if we use the CHY recipe for both gauge and gravity amplitudes.
\subsection{Summary}
Now we summarize what we have done in this section. The main conclusion is that the special solution (\ref{eq:sl1}) supports general MHV gauge and gravitational amplitudes. If we permute the negative helicity particles around, \eqref{eq:sl1} will always return the correct amplitudes, without the help of other solutions.

An immediate question one may ask is what roles do the other $(n-3)!-1$ solutions play in the MHV case? Actually, it has been proposed in \cite{Cachazo:2016sdc} that all the other solutions do not contribute to the MHV amplitudes. Using the machinery worked out in this section, we can give an algebraic characterization between \eqref{eq:sl1} and the other solutions. This is the main subject of the next section.
\section{Other solutions at MHV and Non-MHV}
\label{Sec4}
First, we note that similar result as in Sec.~\ref{Sec3} can also be proved for anti-MHV amplitudes using the other special solution (\ref{eq:sl2}). All we need to do is to exchange angular and square brackets.

As to other solutions at MHV, it is not difficult to check numerically (we have checked up to $9$-point) that if we plug any solution other than \eqref{eq:sl1} into \eqref{generalp1}, we get $\text{Pf}\,'(\Psi)=0$, due to the fact that
\begin{equation}
    \det\left(C^{x,y,m}_{x,j,n}\right)=0\,.\Label{detC=0}
\end{equation}
However, we can study this problem from another direction, namely, we can solve from the independent set of \eqref{detC=0} all the solutions to the scattering equation except for the special one (\ref{eq:sl1}). In other words, if we put together \eqref{detC=0} and the scattering equation (\ref{eq:se}), the solution set will be those of \eqref{eq:se} that do not contribute to the MHV amplitudes. We hope that this is the first step towards understanding and classifying the Eulerian number pattern of the solution set \cite{Cachazo:2016sdc}. In the following, we call the solution (\ref{eq:sl1}) the \textit{MHV solution} and (\ref{eq:sl2}) the \textit{anti-MHV solution}. All the others are thus called \textit{non-MHV solutions}, since they only contribute to certain non-MHV amplitudes.


\subsection{Independent set of characteristic equations}
Once we change $x$, $y$, $m$ and $j$ in \eqref{detC=0}, we get a new equation that should be satisfied by the non-MHV solutions. However, such a system of equations is redundant, out of which we need to extract a complete and independent set.

For a given $x$, which is the position of the first negative helicity particle, and the gauge choice (\ref{xygauge}), the entries of matrix $C$ can be written as
\begin{align}
    C_{ab}&=-\sqrt{2}\frac{[ab]\langle xb\rangle}{\sigma_{ab}\langle xa\rangle}\,,& &1\leq a\neq x\leq n-1\text{ and }b\neq a\,. \nonumber\\
    C_{aa}&=\sqrt{2}\sum_{\substack{l=1\\l\neq a}}^{n}\frac{[al]\langle xl\rangle}{\sigma_{al}\langle xa\rangle}\,,& &1\leq a\leq n\,.
\end{align}
After extracting common factors of each line and column, we get
\begin{equation}
\label{eq:detD0}
    \det\left(C^{x,y,m}_{x,j,n}\right)=(\sqrt{2})^{n-3}\frac{\langle xy\rangle\langle mx\rangle}{\langle xj\rangle\langle nx\rangle}\det\left(D^{x,y,m}_{x,j,n}\right),
\end{equation}
where
\begin{align}
    &D_{ab}=-\frac{[ab]}{\sigma_{ab}}\,,& &D_{aa}=\sum_{\substack{l=1\\l\neq a}}^{n}\frac{[al]\langle xl\rangle}{\sigma_{al}\langle xa\rangle}\,.
\end{align}
The range of the indices in $D$ is the same as that of $C$. We find that the following quantity only depends on $x$
\begin{equation}
\label{Dx}
    \mathscr{D}_{x}\equiv\frac{\text{perm}(xym)\text{perm}(xjk)}{\langle xy\rangle\sigma_{ym}\langle mx\rangle\langle xj\rangle\sigma_{jk}\langle kx\rangle}\det\left(D_{x,j,k}^{x,y,m}\right).
\end{equation}
The proof is very similar to that of $\text{Pf}\,'(\Psi)$ be independent of $i$ and $j$ given in \cite{Cachazo:2013hca}. In terms of this new quantity, \eqref{generalp1} can be rewritten as
\begin{equation}
    \text{Pf}\,'(\Psi)=(-1)^{s(n)}(\sqrt{2})^{n}\langle xy\rangle^{4}\mathscr{D}_{x}C_{xx}\,.
\end{equation}
Next, we show that there is no $\sigma_{x}$ contained in $\mathscr{D}_{x}$. Indeed, since both the row-$(x)$ and column-$(x)$ is deleted, the only place $\sigma_{x}$ can appear is in the diagonal entry $D_{aa}$, as $1/\sigma_{ax}$. However, because of the $\langle xl\rangle$ in the numerator, the coefficient of $1/\sigma_{ax}$ is actually zero. Consequently
\begin{align}
    &\mathscr{D}_{x}=0& &(1\leq x\leq n-3)\Label{Dx=0}
\end{align}
forms a complete and independent system of polynomial equations for our $n-3$ unknown $\sigma$'s. For $n=5$, there is only one solution to \eqref{Dx=0}, which is exactly \eqref{eq:sl2}. We have numerically studied a number of cases up to $n=9$, and find that the $(n-3)!-1$ non-MHV solutions of the scattering equation all satisfy \eqref{Dx=0}. On the other hand, \eqref{Dx=0} contains additional solutions other than those of the scattering equation. We have confirmed this fact numerically at $n=6$.

As an example, it is easy to show analytically that the anti-MHV solution (\ref{eq:sl2}) indeed satisfies $\mathscr{D}_{x}=0$ for arbitrary $n$. In this case, we have
\begin{align}
    &D_{ab}=-\frac{[a\rchi][b\rchi][n-1,n-2]}{[n-2,\rchi][n-1,\rchi]}\,,& &D_{aa}=-\frac{[a\rchi]^{2}[n-1,n-2]}{[n-2,\rchi][n-1,\rchi]}\,,
\end{align}
for $1\leq a\neq b\neq x\leq n$. After we delete two more rows and columns, and extract $[a\rchi]$ from row-$(a)$ and $[b\rchi]$ from row-$(b)$, we will get a matrix whose entries are identical, and it must have zero determinant. Thus we have proved that the solution (\ref{eq:sl2}) leads to $\mathscr{D}_{x}=0$ and makes no contribution to MHV amplitudes. For the other $(n-3)!-2$ non-MHV solutions, at this moment one can use only numerical methods, since no analytic expression for any of them is known in the literature.

Geometrically, the scattering equation (\ref{eq:se}) represents a set of $(n-3)$ hyper-surfaces in the space of $n-3$ complex variables (locally $\mathbb{C}^{n-3}$) while the $(n-3)!$ solutions are just the intersection points of these hyper-surfaces. Meanwhile, \eqref{Dx=0} defines another set of $(n-3)$ hyper-surfaces and their intersection points always have $(n-3)!-1$ in common with the ones given by \eqref{eq:se}. The algebraic geometric property of these two sets of equations needs to be further studied.


\subsection{Non-MHV solutions and Non-MHV amplitudes}
It has been indicated in \cite{Cachazo:2016sdc} that there is an Eulerian number partition pattern in the $(n-3)!$ solutions. Namely, $\text{N}^{k}$MHV amplitudes are only supported by $A(n-3,k)$ solutions, where $A(n-3,k)$ is the $k$-th Eulerian number of index $n-3$. We have also numerically checked this fact up to $n=9$. Our approach described above in Sec.~\ref{Sec3} may potentially be generalized to non-MHV amplitudes, despite the fact that there is no compact analytic expression known for any of the other non-MHV solutions: Working out similar characteristic equations for non-MHV solutions is still promising. We leave this to our future work.

\section{Conclusions and discussions}
\label{Sec5}
In this paper, we have proved in the CHY formalism that the special rational solution (\ref{eq:sl1}) of the scattering equations leads to the Parke-Taylor formula for MHV Yang-Mills amplitudes with an {\em arbitrary} number of external gluons, as well as the Hodges formula for MHV gravity amplitudes. This is achieved by developing techniques to compute relevant reduced Pfaffians/determinants in a manifestly M\"obius covariant formalism. Two useful properties have been introduced and proved, which make the M\"obius invariance of the formalism manifest. Then the fact that another known special solution (\ref{eq:sl2}) supports only the anti-MHV amplitudes follows immediately. By numerical check, we pointed out that all other solutions of the scattering equations do not contribute to the MHV amplitudes at all. Moreover, algebraic conditions satisfied by the $(n-3)!-1$ non-MHV solutions, which do not contribute to the MHV amplitudes, have been established. We leave further study on amplitudes beyond MHV and anti-MHV to future work. 

The correspondence we have established in this paper between the MHV solutions of scattering equations and the MHV Yang-Mills/gravity amplitudes has a profound physical implication: Namely in the CHY formalism for Yang-Mills and gravity, the solutions of scattering equations, involving only external momenta, know mysteriously about the external helicity/polarization configurations of the scattering amplitudes.      

Recently, Lam and Yao developed a systematic method  \cite{Lam:2015sqb,Lam:2016tlk} to evaluate CHY integrations for  $n$-point amplitudes. Their method can be applied to any fixed helicity configuration with manifest M\"obius invariance. Many examples with a small value of $n$ were explicitly calculated in \cite{Lam:2015sqb}, with results in the MHV cases consistent with the Parke-Taylor formula. However, it is not easy to see the correspondence between solutions of scattering equations and helicity configurations in their way. Nevertheless, this work initiates a new approach to study the CHY formula, especially for non-MHV or non-anti-MHV configurations, for which no analytic solution of scattering equations is known. The connection between the approach by Lam and Yao \cite{Lam:2015sqb,Lam:2016tlk} and the current work is an interesting topic and deserves further study.

\acknowledgments
YD would like to acknowledge National Natural Science Foundation of
China under Grant Nos. 11105118, 111547310 and the starting grant of Wuhan University under No. 410100046. We would like to thank the referee for helpful comments, especially for reminding us of the recent work by Lam and Yao.

\appendix
\section{Spinor-helicity formalism}\label{appA}
In this section, we briefly introduce the spinor helicity formalism \cite{Xu:1986xb} and show the conventions we have employed in our calculation. The metric we use is $g_{\mu\nu}=(1,-1,-1,-1)$. The usual Dirac $u$ and $v$ spinor can be defined as
\begin{equation}
u^{s}(p)=\left(\begin{array}{c}
\xi_{\alpha}(p,s) \\ \eta^{\dagger\dot{\alpha}}(p,s) \\
\end{array}\right),\quad v^{s}(p)=\left(\begin{array}{c}
\eta_{\alpha}(p,s) \\ \xi^{\dagger\dot{\alpha}}(p,s) \\
\end{array}\right)\,,
\end{equation}
such that the Dirac conjugate is:
\begin{align}
&\overline{u}^{s}(p)=\left(\;\eta^{\alpha}(p,s),\;\xi^{\dagger}_{\dot{\alpha}}(p,s)\;\right)\,,& &\overline{v}^{s}(p)=\left(\;\xi^{\alpha}(p,s),\;\eta^{\dagger}_{\dot{\alpha}}(p,s)\;\right)\,.
\end{align}
Here $\xi$ and $\eta$ are Weyl spinors. The dotted and undotted indices are converted through the conjugation $\dagger$ while they are raised and lowered by:
\begin{equation}
\epsilon^{12}=-\epsilon^{21}=1,\quad\epsilon_{12}=-\epsilon_{21}=-1\,.
\end{equation}
$u$ and $v$ satisfy the Dirac equation:
\begin{align}
&\left(\gamma^{\mu}p_{\mu}-m\right)u^{s}(p)\equiv\left(\slashed{p}-m\right)u^{s}(p)=0\,, & \overline{u}^{s}(p)\left(\slashed{p}-m\right)=0\,,\nonumber\\
&\left(\gamma^{\mu}p_{\mu}+m\right)v^{s}(p)\equiv\left(\slashed{p}+m\right)v^{s}(p)=0\,, & \overline{v}^{s}(p)\left(\slashed{p}+m\right)=0\,.
\end{align}
In the massless limit, $s$ labels the helicity and we have the special solution:
\begin{align}
\label{eq:xyspinor}
&\xi_{\alpha}(p,+)=0\,, && \xi_{\alpha}(p,-)=\sqrt{2E}\left(\begin{array}{c}
-e^{-i\phi/2}\sin\frac{\theta}{2}  \\ e^{i\phi/2}\cos\frac{\theta}{2} \\
\end{array}\right)\,,\nonumber\\
&\eta_{\alpha}(p,+)=\sqrt{2E}\left(\begin{array}{c}
-e^{-i\phi/2}\sin\frac{\theta}{2}  \\ e^{i\phi/2}\cos\frac{\theta}{2} \\
\end{array}\right)\,, && \eta_{\alpha}(p,-)=0\,,
\end{align}
for the momentum $p^{\mu}=(E,E\sin\theta\cos\phi,E\sin\theta\sin\phi,E\cos\theta)$. Other nonzero two-spinors are related to them by
\begin{align}
&\xi^{\dagger\dot{\alpha}}(p,-)=\eta^{\dagger\dot{\alpha}}(p,+)=\sqrt{2E}\left(\begin{array}{c}
e^{-i\phi/2}\cos\frac{\theta}{2} \\ e^{i\phi/2}\sin\frac{\theta}{2} \\
\end{array}\right)\,,\nonumber\\*
&\xi^{\alpha}(p,-)=\eta^{\alpha}(p,+)=\sqrt{2E}\left(\;e^{i\phi/2}\cos\frac{\theta}{2},\;e^{-i\phi/2}\sin\frac{\theta}{2}\;\right)\,,\nonumber\\*
&\xi^{\dagger}_{\dot{\alpha}}(p,-)=\eta^{\dagger}_{\dot{\alpha}}(p,+)=\sqrt{2E}\left(\;-e^{i\phi/2}\sin\frac{\theta}{2},\; e^{-i\phi/2}\cos\frac{\theta}{2}\;\right)\,.
\end{align}
The normalization $\sqrt{2E}$ agrees with the one used in \cite{Elvang:2013cua}. We define the new angular and square bracket notation for the spinors:
\begin{align}
&u^{+}(p)=v^{-}(p)=\left(\begin{array}{c}
0 \\ \eta^{\dagger\dot{\alpha}}(p,+) \\
\end{array}\right)\equiv\left|p\right\rangle\,,& &u^{-}(p)=v^{+}(p)=\left(\begin{array}{c}
\xi_{\alpha}(p,-) \\ 0 \\
\end{array}\right)\equiv\left|p\right]\,,\nonumber\\
&\overline{u}^{+}(p)=\overline{v}^{-}(p)=\Big(\;\eta^{\alpha}(p,+),\;0\;\Big)\equiv\left[p\right|\,,& &\overline{u}^{-}(p)=\overline{v}^{+}(p)=\left(\;0,\;\xi^{\dagger}_{\dot{\alpha}}(p,-)\;\right)\equiv\left\langle p\right|\,.
\end{align}
Then using this notation, a light like four-vector can be expressed in terms of the spinors as
\begin{align}
    &p^{\mu}=\frac{1}{2}\left\langle p\left|\gamma^{\mu}\right|p\right]=\frac{1}{2}\left[p\left|\gamma^{\mu}\right|p\right\rangle\,,& &\slashed{p}=\left|p\left\rangle\right[p\right|+\left|p\left]\right\langle p\right|\,,
\end{align}
and the Mandelstam variable $s_{ij}$ can be expressed as
\begin{equation}
    s_{ij}=2p_{i}\cdot p_{j}=-\langle p_{i}p_{j}\rangle[p_{i}p_{j}]\,.
\end{equation}
\section{Proof of \eqref{Property-1} and \eqref{Property-2}}\label{appB}
In the following subsections, we prove the two properties given in \eqref{Property-1} and \eqref{Property-2}. In the derivation, if we say, for example, $B$ part of a matrix, we mean those entries that belong to the \textit{original} $B$ sub-matrix in $\Psi$.

\subsection{Proof of property-1}\label{sec:p1}
\begin{figure}[t]
\centering
	\includegraphics[width=0.45\textwidth]{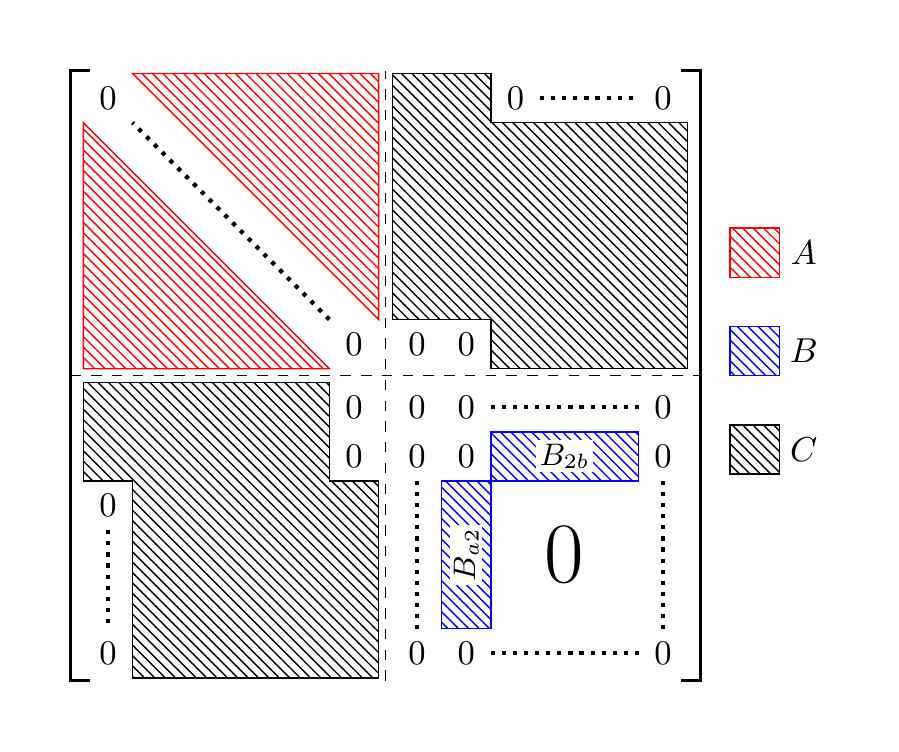}
	\caption{\label{fig:Psi} The structure of the matrix $\Psi$ after we fixed the gauge. Only the shaded regions are generally nonzero.}
\end{figure}

Now let us prove \eqref{Property-1} by recursively expanding the Pfaffian using the formula
\begin{equation}
    \text{Pf}(X)=\sum_{\substack{j=1\\j\neq i}}^{2N}(-1)^{i+j+1+\theta(i-j)}x_{ij}\text{Pf}(X_{ij}^{ij})
\end{equation}
for a $2N\times 2N$ anti-symmetric matrix $X=(x_{ij})$. The proof consists of the following steps:
\begin{enumerate}[label=\bf \emph{(\roman*)}]
\item {\bf \textit{The structure of $\Psi$}}\label{step1}

As our first step, we substitute the polarizations given by \eqref{Eq:Gauge} into the $B$ and $C$ matrices defined by \eqref{Eq:ABCMatrices}.
Under the choice of reference momenta, the only nonzero $\epsilon_i\cdot\epsilon_j$ in the $B$ matrix are ${\epsilon_{2}^{-}\cdot\epsilon_{b}^{+}}$ with $3\leq b\leq n-1$. Thus only the second row and the second column contain nonzero entries. Under our chioce of gauge \eqref{Eq:Gauge}, we also have
\bea
    \epsilon^{-}_{1,2}\cdot k_{n}^{}=0,~~~~\epsilon_{a}^{+}\cdot k_{1}^{}=0~~~~(3\leq a\leq n).
\eea
Then the last $n-2$ entries of the first column (row) as well as the first two entries of the last column (row) in $C$ ($-C^T$) matrix have to be zero. Hence the general structure of the matrix $\Psi$ has the form shown by \figref{fig:Psi}.

\item {\bf \textit{The expansion of} $\text{Pf}\,'(\Psi)$}

To calculate $\text{Pf}\,'(\Psi)$, we choose to delete the $(n-1)$-th and $n$-th row and column, which leads to
\begin{equation}
    \text{Pf}\,'(\Psi)=\frac{-1}{\sigma_{n-1,n}}\text{Pf}\left(\Psi_{n-1,n}^{n-1,n}\right)\equiv\frac{-1}{\sigma_{n-1,n}}\text{Pf}(\widetilde{\Psi})\,.\Label{Eq:ReducedPfaffian}
\end{equation}
We now expand $\text{Pf}(\widetilde{\Psi})$  with respect its $n$-th row, which is row-$(2)$ of $B$ and $C$, and obtain
\begin{equation}
    \text{Pf}(\widetilde{\Psi})=\sum_{\substack{b=1\\b\neq n}}^{2n-2}(-1)^{n+b+1+\theta(n-b)}\widetilde{\Psi}_{nb}\text{Pf}\left(\widetilde{\Psi}_{n,b}^{n,b}\right)\,,\Label{Eq:Pfaffian1}
\end{equation}
where
\begin{equation}
    \widetilde{\Psi}_{nb}=\left\{\begin{array}{l @{\hspace{1em}} l}
        C_{2b} & 1\leq b\leq n-2 \\
        B_{2,b-n+2} & n-1\leq b\leq 2n-2
    \end{array}\right..
\end{equation}

\item {\bf \textit{The reduction of} $\text{Pf}\left(\widetilde{\Psi}_{n,b}^{n,b}\right)$ \textit{in \eqref{Eq:Pfaffian1}}}

It is not difficult to see that all the sub-Pfaffians $\text{Pf}\left(\widetilde{\Psi}_{n,b}^{n,b}\right)$ with $1\leq b\leq n-2$ are zero. In this case the $\widetilde{\Psi}_{n,b}^{n,b}$ has a zero $B$ part, which is still $n\times n$ dimensional. However, the nonzero off-diagonal $C$ part has dimension $(n-1)\times(n-3)$, as illustrated in Fig.~\ref{fig:subpsi1}. Then by elementary transformations, we can always make two rows of the $C$ part zero, such that we get a matrix with two entire rows zero, which has zero determinant. Since elementary transformations do not change determinant, we must have
\begin{figure}[t]
    \centering
\subfloat[$1\leq b\leq n-2$]{
\includegraphics[width=0.4\textwidth]{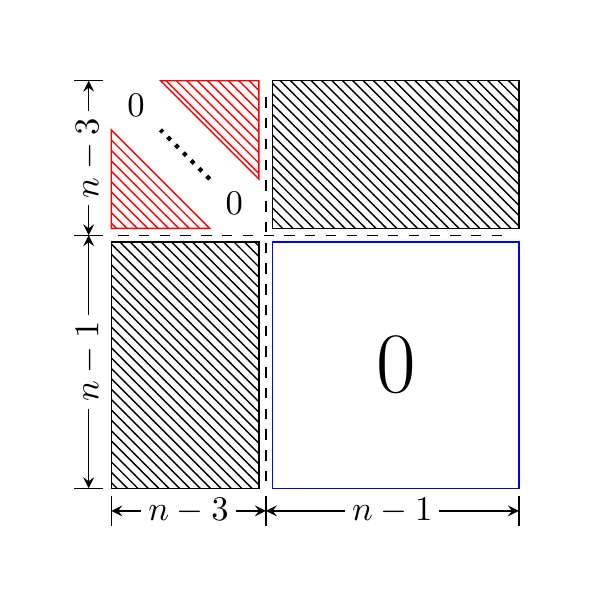}
\label{fig:subpsi1}
}
\qquad\qquad
\subfloat[$n-1\leq b\leq 2n-2$]{
\includegraphics[width=0.4\textwidth]{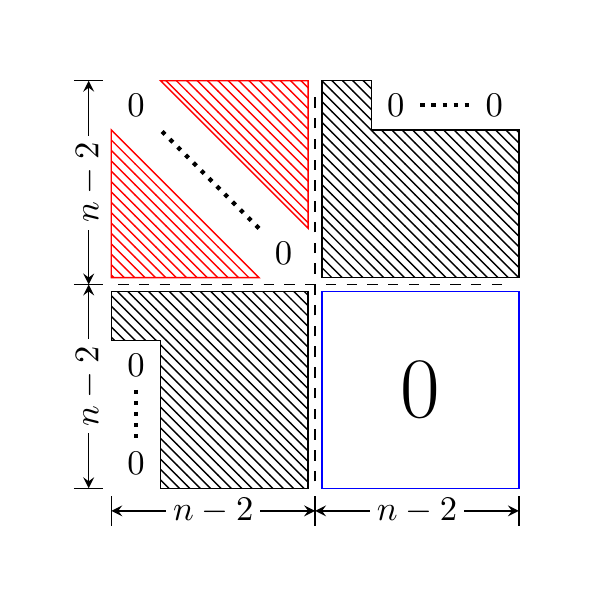}
\label{fig:subpsi2}
}
\caption{(a) The submatrix of $\widetilde{\Psi}$ when the deleted column is in the $C$ part during the recursive expansion of $\text{Pf}(\widetilde{\Psi})$. The Pfaffian of this submatrix is zero. (b) The structure of $\psi_{m}$, which has nonzero Pfaffian in general. It is the submatrix obtained when the deleted column is in the $B$ part during the recursive expansion of $\text{Pf}(\widetilde{\Psi})$. }
\end{figure}
\bea
   \text{Pf}\left(\widetilde{\Psi}_{n,b}^{n,b}\right)=\sqrt{\det\left(\widetilde{\Psi}_{n,b}^{n,b}\right)}=0~~~~(1\leq b\leq n-2).~~\Label{zeroPf}
\eea
We then re-express $\text{Pf}(\widetilde{\Psi})$ by
\begin{equation}
    \text{Pf}(\widetilde{\Psi})=\sum_{m=3}^{n}(-1)^{b+1}B_{2m}\text{Pf}\left(\widetilde{\Psi}_{n,m+n-2}^{n,m+n-2}\right)\equiv\sum_{m=3}^{n}(-1)^{m+1}B_{2m}\text{Pf}\left(\psi_{m}\right).\Label{Eq:Pfaffian2}
\end{equation}
The summation starts from $m=3$ since $B_{21}\sim \epsilon_2^-\cdot\epsilon_1^-$ vanishes due to our choice of gauge and $B_{22}=0$ by definition. The general structure of $\psi_{m}$ thus can be shown by \figref{fig:subpsi2}.

\item {\bf \textit{The reduction of} $\text{Pf}\left(\psi_{m}\right)$ \textit{in \eqref{Eq:Pfaffian2}}}

We apply the reduction process in {(iii)} on $\psi_{m}$, and expand it with respect to its $(n-1)$-th row, which is row-$(1)$ of $C$
\begin{equation}
    \text{Pf}(\psi_{m})=\sum_{s=1}^{n-2}(-1)^{n+s+1}C_{1s}\text{Pf}\left([\psi_{m}]_{n-1,s}^{n-1,s}\right)\,.\Label{Eq:Pfaffian3}
\end{equation}
If $s\neq 1$, the Pfaffian of the corresponding submatrix of $\psi_{m}$ is zero. The proof is similar to the previous one. In this case, the first row and column of $\psi_{m}$ is inherited by the submatrix. Both the first row and first column are entirely zero in the $C$ part. Then we can exchange both row-$(1)$ and column-$(1)$ with row-$(n-2)$ and column-$(n-2)$ in the $A$ part. This manipulation only changes the sign of the Pfaffian in question. The resultant matrix has a $(n-2)\times(n-2)$ dimensional zero block at the bottom right corner, while the off-diagonal nonzero block is $(n-2)\times(n-4)$ dimensional. Thus the Pfaffian is zero for the same reason as given in the paragraph above \eqref{zeroPf}. This process is illustrated in Fig.~\ref{fig:psim}.
\begin{figure}[t]
\centering
\includegraphics[width=0.7\textwidth]{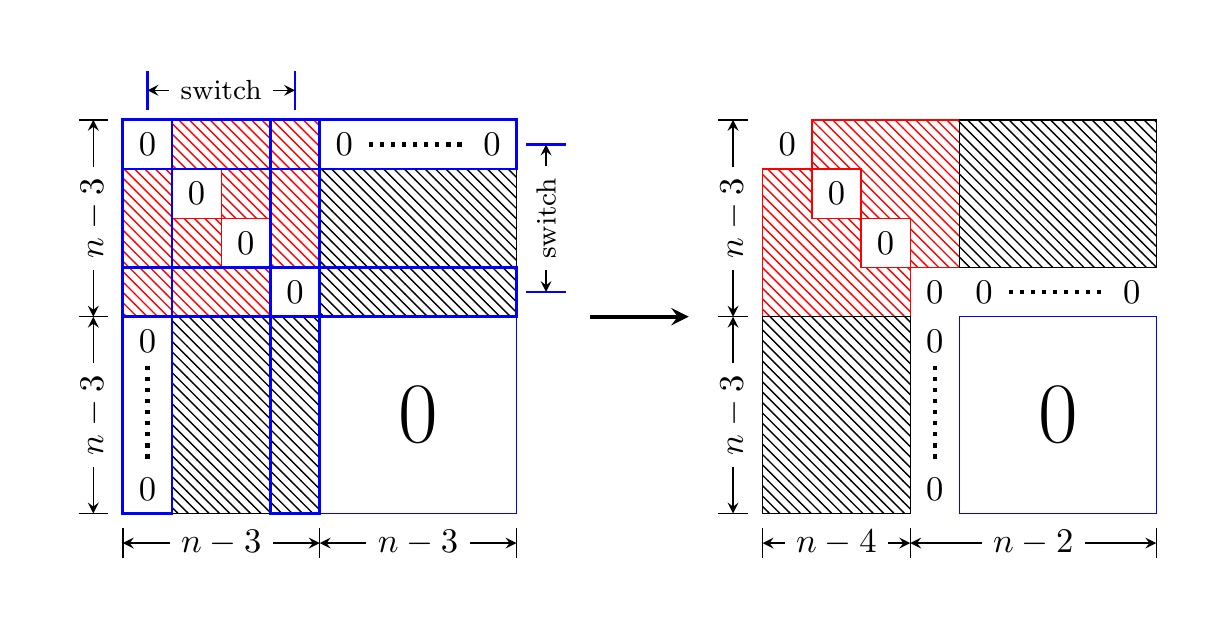}
\caption{\label{fig:psim} The $(2n-6)\times(2n-6)$ dimensional submatrix of $\psi_{m}$ when the deleted column is not the first one. The Pfaffian of this submatrix is zero, which can be told from switching two rows and columns}
\end{figure}
Therefore we have only one sub-Pfaffian contributing
\begin{equation}
    \text{Pf}(\psi_{m})=(-1)^{n}C_{11}\text{Pf}\left([\psi_{m}]_{n-1,1}^{n-1,1}\right)\equiv(-1)^{n}C_{11}\text{Pf}(\psi'_{m})\,.\Label{Eq:Pfaffian4}
\end{equation}

\item {\bf \textit{The expression of} $\text{Pf}(\psi'_{m})$ \textit{in \eqref{Eq:Pfaffian4}}}\label{step5}

It is now instructive to see what the $(2n-6)\times(2n-6)$ dimensional matrix $\psi'_{m}$ is composed of in terms of the original $A$, $B$ and $C$ matrix. It is straightforward to observe that
\begin{equation}
\renewcommand{\arraystretch}{2.5}
    \psi'_{m}=\left(\begin{array}{>{\displaystyle}c:>{\displaystyle}c}
    A_{1,n-1,n}^{1,n-1,n} & -\left(C^{1,2,m}_{1,n-1,n}\right)^{T} \\ \hdashline
    C^{1,2,m}_{1,n-1,n} & 0
    \end{array}\right)\,.
\end{equation}
In the off-diagonal blocks of $\psi'_{m}$, we have the original matrix $C$ with row-$(1,2,m)$ and column-$(1,n-1,n)$ deleted. If the determinant of $C_{1,2,m}^{1,n-1,n}$ is zero, then $\text{Pf}(\psi'_{m})=0$ since elementary transformations can make one entire row in the lower half of $\psi'_{m}$ zero. If $C_{1,2,m}^{1,n-1,n}$ has a nonzero determinant, we  always find an elementary transformation that makes the $A$ part of $\psi'_{m}$ zero. For example, to make the first row of the $A$ part zero, we need to find $x_{3},\cdots, x_{m-1},{x}_{m+1},\cdots, x_{n}$ from the following set of equations
\begin{gather}
    x_{3}C_{3l}+x_{4}C_{4l}+\cdots+x_{m-1}C_{m-1,l}+x_{m+1}C_{m+1,l}+\cdots+x_{n}C_{nl}=-A_{2l}\,,\nonumber\\
    (2\leq l\leq n-2)\,.
\end{gather}
Now that we have $n-3$ unknowns with $n-3$ equations, we can always find a solution when $C^{1,2,m}_{1,n-1,n}$ has nonzero determinant. Then multiplying each row in $C^{1,2,m}_{1,n-1,n}$ by the corresponding $x$, adding it to the first row of $A$, we can make the first row of the $A$ part zero. Continuing this operation to all rows of the $A$ part, we can thus make the entire block zero. Therefore, both situations can be captured in the following equation
\begin{equation}
\renewcommand{\arraystretch}{2.5}
    \text{Pf}(\psi'_{m})=\text{Pf}\left(\begin{array}{>{\displaystyle}c:>{\displaystyle}c}
    0 & -\left(C^{1,2,m}_{1,n-1,n}\right)^{T} \\ \hdashline
    C^{1,2,m}_{1,n-1,n} & 0
    \end{array}\right)=(-1)^{\frac{(n-2)(n-3)}{2}}\det\left(C^{1,2,m}_{1,n-1,n}\right).~~\Label{Eq:Pfaffian5}
\end{equation}
Now if we put together \eqref{Eq:ReducedPfaffian}, \eqref{Eq:Pfaffian2}, \eqref{Eq:Pfaffian4} and \eqref{Eq:Pfaffian5}, we  get \eqref{Property-1}, and the the proof of Property-$1$ is complete.
\end{enumerate}
\paragraph{Feynman diagram analysis.}\hspace{-1em}\footnote{Similar analysis can be found in e.g., \cite{Elvang:2013cua}.}
Although \eqref{Property-1} have been derived from properties of Pfaffians structures, it is worth pointing out that the reductions \ref{step1}-\ref{step5} can be understood more physically from Feynman diagrams. In the amplitudes calculated from usual Feynman diagrams, polarization for any external gluon must be contracted with either another polarization or an external momentum. In $n$-gluon tree diagrams, the number of polarizations is $n$ and the number of vertices in a Feynman diagram should be at most $n-2$. Thus we at least have one factor of $\epsilon_i\cdot\epsilon_j$. As already mentioned in \ref{step1}, the nonzero $\epsilon_i\cdot\epsilon_j$ can only be $\epsilon^-_2\cdot\epsilon^+_b$ $(3\leq b\leq n-1)$. Since the polarization $\epsilon^-_2$ can only appear once in one diagram, we only have one nonzero factor of the type $\epsilon_i\cdot\epsilon_j$ for each Feynman diagram. Thus MHV partial amplitude can be written as a summation of the terms proportional to $\epsilon^-_2\cdot\epsilon^+_b$, which agrees with \eqref{Eq:Pfaffian2}. Meanwhile, all the other $n-2$ polarizations have to be contracted with $n-2$ external momenta. Next we study whether we can have $k_i\cdot k_j$ in the summand. The most possible contributing diagrams are those constructed by only cubic vertices, each of which contributes a factor of the form $k^{\mu}\eta^{\rho\sigma}$ to each summand. An $n$-gluon tree diagram at most contains $n-2$ vertices such that the vertices contribute $3(n-2)$ Lorentz indices. However, since we have $2(n-3)$ propagators contracting with vertices, thus the total number of external Lorentz indices is $n$. Then the $n$ external polarizations can either contract with $k^{\mu}$ or $\eta^{\rho\sigma}$.
\begin{itemize}
\item If all the $(n-2)$ momenta $k$ contract with external polarizations (i.e., there is no $k_i\cdot k_j$), there must be two polarizations left and have to contract with each other. This case is allowed because we do have nonzero contractions $\epsilon_2\cdot\epsilon_b$ $(3\leq b\leq n-1)$ available.

\item If there exists a factor $k_i\cdot k_j$, we should have at least two less $k$'s contract with polarizations. Thus we have at least two more polarizations contract with each other via $\eta^{\rho\sigma}$, which should vanish since there are no more nonzero $\epsilon_i\cdot\epsilon_j$.
\end{itemize}
Thus for MHV case with our gauge choice (\ref{Eq:Gauge}), Feyman diagram can only contribute term that contains just one signle factor $(\epsilon^-_2\cdot\epsilon_b)$ $(2<b\leq n-1)$ and $(n-2)$ factors of the type $\epsilon\cdot k$. In the CHY language, it means that $\text{Pf}\,'(\Psi)$ should not contain any entries in the $A$ matrix, which also agrees with \eqref{Eq:Pfaffian3}.

\subsection{Proof of property-2}\label{sec:p2}
We now turn to prove the three relations in \eqref{Property-2}.

\paragraph{\bf{The entries of $\Phi$.}}

We plug in \eqref{eq:sl1} and write $\Phi$ in terms of spinor products. Using \eqref{Eq:Sigmaij}, we  express the entries of $\Phi$ as
\begin{align}
    \Phi_{ab}&=\frac{s_{ab}}{\sigma_{ab}^{2}}=-\frac{[ab]\langle a\rchi\rangle^{2}\langle b\rchi\rangle^{2}\langle n-1,n-2\rangle^{2}}{\langle ab\rangle\langle n-2,\rchi\rangle^{2}\langle n-1,\rchi\rangle^{2}}& &(a\neq b)\,,\Label{Psi1}\\
    \Phi_{aa}&=-\sum_{l\neq a}^{n}\Phi_{al}=\frac{\langle a\rchi\rangle^{2}\langle n-1,n-2\rangle^{2}}{\langle n-2,\rchi\rangle^{2}\langle n-1,\rchi\rangle^{2}}\sum_{l\neq a}^{n}\frac{[al]\langle l\rchi\rangle^{2}}{\langle al\rangle}& &(\text{diagonal})\,.\Label{Psi2}
\end{align}

\paragraph{\bf{The determinant of $\Phi^{1,2,m}_{1,n-1,n}$ with $3\leq m\leq n-1$.}}
Now we calculate $\det\left(\Phi^{1,2,m}_{1,n-1,n}\right)$ with row-$(1,2,m)$ and column-$(1,n-1,n)$ removed.
\begin{itemize}
\item From \eqref{Psi1} and \eqref{Psi2}, we know that each entry of the $\Phi$ matrix contains a factor $F^{2}_{\chi}$ such that We should thus have in all $n-3$ rows containing this common factor. By extracting all of them out of the determinant, we get an overall factor
\bea
 \left(F_\chi\right)^{2(n-3)}.~~\Label{Factor1}
\eea
\item As shown by \eqref{Psi1}, each $\Phi_{ab}$ $(a\neq b)$ contains a factor $\Spaa{a\rchi}^2\Spaa{b\rchi}^2$. It thus is tempting to extract $\langle a\rchi\rangle^{2}$ out of each row and $\Spaa{b\rchi}^{2}$ out of each column, but the obstacle is in $\Phi_{aa}$, which seem to contain only $\Spaa{a\rchi}^{2}$ instead of $\Spaa{a\rchi}^{4}$, as in \eqref{Psi2}. Now let us show that a $\Phi_{aa}$ $(a\neq n)$ secretly contains one more $\Spaa{a\rchi}^2$ if we apply the Schouten identity properly. First, we define the following quantities for convenience
\begin{align}
    &P(a)\equiv\prod_{\substack{c=1\\c\neq a}}^{n}\langle ac\rangle\,,& &P(a,l)\equiv\prod_{\substack{c=1\\c\neq a,l}}^{n}\langle ac\rangle\,.
\end{align}
Then we  rewrite the summation in $\Phi_{aa}$ as
\begin{align}
    \sum_{\substack{l=1\\l\neq a}}^{n}\frac{[al]\langle l\rchi\rangle^{2}}{\langle al\rangle}=\frac{1}{P(a)}\sum_{\substack{l=1\\l\neq a}}^{n}[al]\langle l\rchi\rangle^{2}P(a,l)\,.
\end{align}
Starting from $l=2$, we always have $\langle a1\rangle$ in $P(a,l)$ (for $a\geq 2$) and we use the Schouten identity
\begin{equation}
    \langle l\rchi\rangle\langle a1\rangle=\langle a\rchi\rangle\langle l1\rangle+\langle al\rangle\langle 1\rchi\rangle
\end{equation}
such that
\begin{equation}
    \sum_{\substack{l=1\\l\neq a}}^{n}[al]\langle l\rchi\rangle^{2}P(a,l)=P(a,1)\langle 1\rchi\rangle\sum_{\substack{l=1\\l\neq a}}^{n}[al]\langle l\chi\rangle+\langle a\rchi\rangle\sum_{\substack{l=1\\l\neq a}}^{n}\frac{[al]\langle l\rchi\rangle\langle l1\rangle}{\langle a1\rangle}P(a,l)\,.
\end{equation}
The first sum yields zero because of momentum conservation, while the second term leads to
\begin{equation}
    \Phi_{aa}=\frac{\langle a\rchi\rangle^{4}\langle n-1,n-2\rangle^{2}}{\langle n-2,\rchi\rangle^{2}\langle n-1,\rchi\rangle^{2}}\sum_{\substack{l=1\\l\neq a}}^{n}\frac{[al]\langle l\rchi\rangle\langle l1\rangle}{\langle al\rangle\langle a\rchi\rangle\langle a1\rangle}\,.
\end{equation}
This is correct for $2\leq a\leq n$, which is adequate for our purpose since the first line has already been deleted. Now we can extract one $\langle an\rangle^{2}$ from each row with $3\leq a\neq m\leq n$, and one $\langle bn\rangle^2$ from each column with $2\leq b\leq n-2$. Then we have a factor
\bea
\left(\prod_{a=1}^{n-1}\langle a\rchi\rangle^{4}\right)\left(\frac{1}{\langle 1\rchi\rangle\langle 2\rchi\rangle\langle m\rchi\rangle\langle 1\rchi\rangle\langle n-1,\rchi\rangle\langle n\rchi\rangle}\right)^{2}\,.\Label{Factor2}
\eea
\end{itemize}
After collecting all these factors in \eqref{Factor1} and \eqref{Factor2}, we  reduce the determinant into the form
\begin{align}
    \det\left(\Phi^{1,2,m}_{1,n-1,n}\right)&=(-1)^{n-3}F_{\chi}^{2n-6}\left(\prod_{a=1}^{n}\langle a\rchi\rangle^{4}\right)\nonumber\\
    &\quad\times\left(\frac{1}{\langle 1\rchi\rangle\langle 2\rchi\rangle\langle m\rchi\rangle\langle 1\rchi\rangle\langle n-1,\rchi\rangle\langle n\rchi\rangle}\right)^{2}\det\left(\phi^{1,2,m}_{1,n-1,n}\right)\,.\Label{eq:detphi}
\end{align}
To settle \eqref{Eq:HodgesMatrix} into a form that is easier to generalize, we define the following two quantities
\begin{align}
    &d_{abc}=d^{abc}\equiv\frac{1}{\langle a\rchi\rangle\langle b\rchi\rangle\langle c\rchi\rangle}\,.
\end{align}
Then \eqref{Eq:HodgesMatrix} becomes
\begin{equation}
    \det\left(\Phi^{1,2,m}_{1,n-1,n}\right)=(-1)^{n-3}\left(F_{\chi}\right)^{2n-6}\left(P_{\chi}\right)^{4}\left(d_{1,2,m}d^{1,n-1,n}\right)^{2}\det\left(\phi_{1,n-1,n}^{1,2,m}\right)\,.\Label{HM2}
\end{equation}
\paragraph{\bf{The reduced determinant ${\det}'(\Phi)$.}}
It is straightforward to find that after using \eqref{eq:sl1}, we have
\begin{equation}
    \sigma_{12}\sigma_{2m}\sigma_{m1}\sigma_{1,n-1}\sigma_{n-1,n}\sigma_{n1}=\left(\frac{1}{F_{\chi}}\right)^{6}\left(c_{1,2,m}c^{1,n-1,n}\right)^{-1}\left(d_{1,2,m}d^{1,n-1,n}\right)^{2}\,.\Label{sigmachain}
\end{equation}
Using \eqref{HM2} and \eqref{sigmachain}, we can arrive at the result given in \eqref{rdPhi}:
\begin{align}
    {\det}\,'(\Phi)=\frac{(-1)^{m+1}}{\sigma_{12}\sigma_{2m}\sigma_{m1}\sigma_{1,n-1}\sigma_{n-1,n}\sigma_{n1}}\det\left(\Phi_{1,n-1,n}^{1,2,m}\right)=\left(F_{\chi}\right)^{2n}\left(P_{\chi}\right)^{4}\bar{M}_n(12\ldots n)\,.
\end{align}
in which $\bar{M}_n$  is given by
\begin{equation}
    \bar{M}_n(12\ldots n)=(-1)^{n+1}(-1)^{m+1}c_{1,2,m}c^{1,n-1,n}\det\left(\phi_{1,n-1,n}^{1,2,m}\right)\,.
\end{equation}
This is just \eqref{Eq:HodgesFormula} with the choice $(i,j,k;p,q,r)=(1,2,m;1,n-1,n)$ such that
\begin{equation*}
    \text{perm}(12m)\,\text{perm}(1,n-1,n)=(-1)^{m}\,.
\end{equation*}
\paragraph{\textbf{The MHV-like factor $\sigma_{12}\sigma_{23}\ldots\sigma_{n1}$.}}
Using \eqref{eq:sl1}, \eqref{Fn} and \eqref{dPn}, it is not difficult to find that
\begin{equation}
    \sigma_{12}\sigma_{23}\ldots\sigma_{n1}=\left(\frac{1}{F_{\chi}}\right)^{n}\frac{D_{n}}{(P_{\chi})^{2}}\,,
\end{equation}
which gives \eqref{MHVfactor}.
\paragraph{\bf{The determinant of $C^{1,2,m}_{1,n-1,n}$.}}
Now we study $\det\left(C^{1,2,m}_{1,n-1,n}\right)$,
where the row-$(1,2,m)$ and the column-$(1,n-1,n)$ in $C$ have been removed. Inserting the polarizations (\ref{Eq:Gauge}) into the  $C$ matrix defined by \eqref{Eq:ABCMatrices}, we get
\begin{align}
    C_{ab}&=-\sqrt{2}\frac{[ab]\langle a\rchi\rangle\langle b\rchi\rangle\langle b1\rangle\langle n-1,n-2\rangle}{\langle ab\rangle\langle a1\rangle\langle n-2,\rchi\rangle\langle n-1,\rchi\rangle}& &(3\leq a\leq n\text{ and }b\neq a)\,,\nonumber\\
    C_{aa}&=\sqrt{2}\frac{\langle a\rchi\rangle^{2}\langle n-1,n-2\rangle}{\langle n-2,\rchi\rangle\langle n-1,\rchi\rangle}\sum_{\substack{l=1\\l\neq a}}^{n-1}\frac{[al]\langle l\rchi\rangle\langle l1\rangle}{\langle al\rangle\langle a\rchi\rangle\langle a1\rangle}& &(3\leq a\leq n-1)\,.
\end{align}
Similar to what we have done to $\Phi$, we  extract $\langle a\rchi\rangle$ from the rows with $3\leq a\neq m\leq n$, $\langle b\rchi\rangle$ from all columns. Then we get:
\begin{align}
    \det\left(C^{1,2,m}_{1,n-1,n}\right)&=(\sqrt{2})^{n-3}\left(F_{\chi}\right)^{n-3}\left(P_{\chi}\right)^{2}d_{1,2,m}d^{1,n-1,n}\det\left(\widetilde{C}^{1,2,m}_{1,n-1,n}\right)\,,\Label{eq:detC1}
\end{align}
in which the matrix $\widetilde{C}$ has entries given by
\begin{align}
    &\widetilde{C}_{ab}=-\frac{[ab]\langle b1\rangle}{\langle ab\rangle\langle a1\rangle}\quad(a\neq b)\,,& &\widetilde{C}_{aa}=\sum_{\substack{l=1\\l\neq a}}^{n-1}\frac{[al]\langle l\rchi\rangle\langle l1\rangle}{\langle al\rangle\langle a\rchi\rangle\langle a1\rangle}\,.
\end{align}
For $3\leq a\leq n$ and $2\leq b\leq n$. In $\widetilde{C}$,  we can extract a common factor $1/\langle 1a\rangle$ from each row, and another common factor $\langle 1b\rangle$ from each column. These two common factors will almost cancel each other outside the determinant while the reminant is
\begin{equation*}
    \frac{\langle 12\rangle\langle 1m\rangle}{\langle 1,n-1\rangle\langle 1n\rangle}
\end{equation*}
due to the mismatch between the range of $a$ and $b$. After doing this, we find that $\widetilde{C}$ reduces to a form that is identical to the Hodges matrix $\phi$ (see \eqref{Eq:HodgesMatrix}). Therefore we have
\begin{equation}
    \det\left(\widetilde{C}^{1,2,m}_{1,n-1,n}\right)=(-1)^{n-3}\frac{\langle 12\rangle\langle 1m\rangle}{\langle 1,n-1\rangle\langle 1n\rangle}\det\left(\phi^{1,2,m}_{1,n-1,n}\right).
\end{equation}
Plugging it into \eqref{eq:detC1}, we get
\begin{equation}
    \frac{-1}{\sigma_{n-1,n}}\det\left(C^{1,2,m}_{1,n-1,n}\right)=-(-\sqrt{2})^{n-3}\left(F_{\chi}\right)^{n-2}\left(P_{\chi}\right)^{2}\left(d_{1,2,m}c^{1,n-1,n}\right)\frac{\langle 12\rangle\langle 1m\rangle}{\langle 1\rchi\rangle}\det\left(\phi^{1,2,m}_{1,n-1,n}\right)\,.\Label{eq:detC}
\end{equation}
Using \eqref{eq:sl1} and \eqref{Eq:Gauge}, we find that
\begin{equation}
    B_{2m}=2F_{\chi}\langle 12\rangle^{2}\left(d_{1,2,m}\right)^{-1}c_{1,2,m}\frac{[mn]}{\langle 1\rchi\rangle[ny]}\,,
\end{equation}
such that $\det\left(\phi^{1,2,m}_{1,n-1,n}\right)$ can be grouped into $\bar{M}_n$, which independent of $m$. Thus we can pull int out and perform the summation over $m$ in \eqref{Property-1}:
\begin{align}
    &\quad\sum_{m=3}^{n}(-1)^{m}B_{2m}\left[\frac{-1}{\sigma_{n-1,n}}\det\left(C_{1,n-1,n}^{1,2,m}\right)\right]\nonumber\\
    &=-(\sqrt{2})^{n-1}\left(F_{\chi}\right)^{n-1}\left(P_{\chi}\right)^{2}\frac{\langle 12\rangle^{3}}{\langle 1\rchi\rangle^{2}}\bar{M}_n(12\ldots n)\sum_{m=3}^{n}\frac{\langle 1m\rangle[mn]}{[n2]}\nonumber\\
    &=-(\sqrt{2})^{n-1}\left(F_{\chi}\right)^{n-1}\left(P_{\chi}\right)^{2}\frac{\langle 12\rangle^{4}}{\langle 1\rchi\rangle^{2}}\bar{M}_n(12\ldots n)\,.
\end{align}
Note that this equation is independent of the gauge choice of the polarizations. This quantity is nothing but the $\mathscr{D}_{1}$ defined in \eqref{Dx} with the special solution (\ref{eq:sl1}) plugged in.
\paragraph{\bf{The reduced Pfaffian $\text{Pf}\,'(\Psi)$.}}
Finally, in \eqref{Property-1} we find that $C_{11}$ is also gauge independent, if \eqref{eq:sl1} is used
\begin{equation}
    C_{11}=-\sqrt{2}F_{\chi}\langle 1\rchi\rangle^{2}.
\end{equation}
As a result, we get
\begin{align}
    \text{Pf}\,'(\Psi)
    &=(-1)^{s(n)}(\sqrt{2})^{n}(F_{\chi})^{n}(P_{\chi})^{2}\langle 12\rangle^{4}\bar{M}_{n}(12\cdots n),
\end{align}
which proves \eqref{reducedPf}. Had we started with another gauge choice of the polarizations, we should arrive at the same result at this point. The gauge invariance is encoded in the property demonstrated in \eqref{gaugeinv}. Now we have completed the proof of all three equations listed in (\ref{Property-2}).

\bibliographystyle{JHEP}
\bibliography{Refs}

\end{document}